\documentclass[11pt]{article}

\usepackage{pgfpages}
\usepackage{amsmath,amssymb,verbatim}
\usepackage{listings}
\usepackage{epsfig}
\usepackage[latin1]{inputenc}

\newcommand{\N}{I\!\!N}

\newcommand{\R}{I\!\!R}

\def\q{\quad}

\begin{document}

\tolerance=5000

\centerline{{\Large{\bf Fractional wave equation and damped waves}}}

\vspace{0.3cm}
\centerline{{\bf Yuri Luchko}}
\vspace{0.3cm}

\centerline{{Department of Mathematics, Physics, and Chemistry}}

\centerline{{Beuth Technical University of Applied Sciences Berlin}}

\centerline{{Luxemburger Str. 10, 13353 Berlin,\ Germany}}

\centerline{{e-mail: luchko@beuth-hochschule.de}}

\vspace{0.2cm}

\begin{abstract}
\noindent
In this paper, a fractional generalization of the  wave equation that describes propagation of damped waves is considered. In contrast to the fractional diffusion-wave equation, the fractional wave equation contains fractional derivatives of the same order $\alpha,\ 1\le \alpha \le 2$ both in space and in time. We show that this feature is a decisive factor for inheriting some crucial characteristics of the wave equation like a constant propagation velocity of both the maximum of its fundamental solution and its gravity and mass centers. Moreover, the first, the second, and the Smith centrovelocities of the damped waves described by the fractional wave equation are constant and depend just on the equation order $\alpha$. The fundamental solution of the fractional wave equation is determined and shown to be a spatial probability density function evolving in time  that possesses finite moments up to the order $\alpha$.  To illustrate analytical findings, results of numerical calculations and numerous plots are presented.

\end{abstract}

\vspace{0.2cm}

\noindent
{\sl MSC 2010}: 26A33, 35C05, 35E05, 35L05, 45K05, 60E99

\noindent
{\sl Key Words}: Caputo fractional derivative, Riesz fractional derivative, fractional wave equation, propagation velocity, gravity center, mass center, centrovelocity, maximum of fundamental solution, damped waves

\section{Introduction}

During the last few decades, fractional order differential equations have been successfully employed for
modeling of many different processes and systems, see e.g. \cite{Uch08} for  different applications of  derivatives and integrals of fractional order in physics, chemistry, 	 engineering, astrophysics, etc. and \cite{Her11} for  applications of fractional differential equations in classical mechanics, quantum
mechanics, nuclear physics, hadron spectroscopy, and quantum field theory. For other interesting models in form of fractional differential equations we refer the reader to \cite{Dub}, \cite{FDL}, \cite{Hil00}-\cite{Luc11_2},
\cite{Mag04}-\cite{Mai_book},  \cite{Met04} to mention only few of many recent publications. 

Among many other applications, models for anomalous transport processes  in form of time- and/or space-fractional advection-diffusion-wave equations enjoyed a particular attention and have been considered by a number of researches since 1980's.  In particular, this kind of phenomena is known to occur  in viscoelastic media that combine characteristics 
of solid-like materials that exhibit waves propagation and fluid-like materials that support diffusion processes (see e.g. the recent book \cite{Mai_book}). It is important to note that anomalous transport models are usually first formulated in stochastic form in terms of the so called continuous time random walk 
processes. The 
time- and/or space-fractional differential equations are then derived from the stochastic models for a special choice of the probability density functions with the 
infinite first or/and second moments (see e.g. \cite{Kla08}, \cite{Luc11_2}, or \cite{Met04}). 

It is well known that whereas diffusion equation describes a process, where a
disturbance of the initial conditions spreads infinitely fast, the propagation velocity of the disturbance is constant for the
wave equation. In a certain sense, the time-fractional diffusion-wave equation of order $\alpha,\ 1<\alpha <2$ interpolates between these two different behaviors: its response to a localized
disturbance spreads infinitely fast, but the maximum of its fundamental solution disperses with a finite velocity $v(t,\alpha)$ that is determined by the formula (see e.g. \cite{Fuj90} or \cite{LucMaiPov})  
\begin{equation}
\label{speed1}
v(t,\alpha) = C_\alpha t^{\frac{\alpha}{2}-1}.
\end{equation}
For $\alpha = 1$ (diffusion), the propagation velocity is equal to zero because of  $C_{1} = 0$, for $\alpha = 2$ (wave propagation) it remains constant and is equal to $C_{2} = 1$, whereas for all intermediate values of $\alpha$ the propagation velocity of the maximum point depends on time $t$ and is a decreasing function that varies from $+\infty$ at time $t=0+$ 
to zero as $t\to +\infty$. This fact makes it difficult to interpret solutions to the fractional diffusion-wave equation as waves in case  $1<\alpha <2$.

In this paper, a fractional wave equation that contains fractional derivatives of the same order $\alpha,\ 1\le \alpha \le 2$ both in space and in time is considered. The fractional derivative in time is interpreted in the Caputo sense whereas the space-fractional derivative is taken in form of the inverse operator to the Riesz potential (Riesz fractional derivative). It turns out that this feature of the fractional wave equation is a decisive factor for inheriting some crucial characteristics of the wave equation. In particular, we show that six different velocities of the damped waves that are described by the fundamental solution of the fractional wave equation (propagation velocity of its maximum, its gravity and mass centers, the first, the second, and the Smith centrovelocities) are constant and depend just on the equation order $\alpha$. 

From the mathematical viewpoint the fractional wave equation we deal with in this paper has been considered in all probability for the first time in \cite{GorIskLuc}, where an explicit formula for the fundamental solution of this equation was derived. In \cite{MaiLucPag}, a space-time fractional diffusion-wave equation with the Riesz-Feller derivative of order $\alpha \in (0,2]$ and skewness $\theta$ has been investigated in detail. A particular case of this equation called neutral-fractional diffusion equation that for $\theta=0$ corresponds to our fractional wave equation  has been shortly mentioned in \cite{MaiLucPag}. Still, to the best knowledge of the author, both in-depth mathematical treatment and physical interpretation of the fractional wave equation seem to be not yet given in the literature. 

The rest of the paper is organized as follows. In the 2nd section,  basic definitions, problem formulation, and some analytical results for an initial-value problem for a model one-dimensional fractional wave equation  are presented. The fundamental solution $G_\alpha$ for this problem is derived in terms of elementary functions for all values of $\alpha,\ 1\le \alpha < 2$. Moreover, $G_\alpha$ is interpreted as a spatial probability density function evolving in time whose moments up to order $\alpha$ are finite. For the fundamental solution $G_\alpha$ both its maximum location and its maximum value are determined in closed form.  Remarkably, the product of the maximum location and the maximum value of $G_\alpha$  is time-independent and just a function of $\alpha$. Finally we show that both the maximum and the gravity and mass centers of the fundamental solution $G_\alpha$ propagate with constant velocities like in the case of the wave equation but in contrast to the wave equation ($\alpha = 2$) these velocities are different each to other for a fixed value of $\alpha,\ 1<\alpha <2$.  Moreover, the first, the second, and the Smith centrovelocities of the damped waves described by the fractional wave equation are shown to be constants that depend just on the equation order $\alpha$. To illustrate analytical findings, results of numerical calculations, numerous plots,  their physical interpretation and discussion are presented in 3rd section. The last section contains some conclusions and open problems for further research.

\section{Analysis of the fractional wave equation}

\subsection{Problem formulation}
In this paper, we deal with the model one-dimensional  fractional wave equation
\begin{equation}
\label{eq}
D^{\alpha}_t u  \, = \,
\frac{\partial^{\alpha}u}{\partial |x|^{\alpha}}, \q x \in \R\,, \q t \in {\R}_+,\ 1\le \alpha \le 2.
\end{equation}
In (\ref{eq}), $u=u(x,t)$ is a real field variable,  $\frac{\partial^{\alpha}}{\partial |x|^{\alpha}}$ is
the Riesz space-fractional
 derivative of order $\alpha $ that is defined below, and
 $D_t^\alpha$	is
 the Caputo time-fractional derivative of order
  $\alpha$: 
\begin{equation}
\label{fd}
(D^{\alpha} f)(t):= (I^{n-\alpha} f^{(n)} )(t), \ n-1<\alpha\le n,\ n\in \N
\end{equation}
$I^\alpha,\ \alpha \ge 0$ being the Riemann-Liouville fractional integral
$$
(I^{\alpha} f)(t):= 
\begin{cases}
\frac{1}{\Gamma(\alpha)} \int_0^{t} (t-\tau)^{\alpha -1} f(\tau)\, d\tau, \  \alpha >0, \\
f(t),\ \alpha =0,
\end{cases}
$$
and $\Gamma$ the Euler gamma function.
For $\alpha=n,\ n\in \N$, the Caputo fractional derivative coincides with the standard derivative of order $n$.  
 
All quantities in (\ref{eq}) are supposed to be dimensionless, so that the coefficient by 
the Riesz space-fractional
 derivative can be taken to be equal to one without loss of generality.  
 
For the equation (\ref{eq}), the initial-value
problem
\begin{equation}
\label{ic}
u(x,0)=   \varphi(x)\,,\ \ \frac{\partial u}{\partial t}(x,0)= 0, \ \ x\in \R
\end{equation}
is considered. In this paper, we are mostly interested in behavior and properties of the 
fundamental solution (Green function) $G_{\alpha}$ 
of  the
equation (\ref{eq}), i.e. its solution  with the initial condition  $\varphi(x) = \delta (x)$, $\delta$ being  the Dirac delta function. 
 
For a sufficiently well-behaved function $f$ the Riesz space-fractional
 derivative of order $\alpha, \ 0<\alpha \le 2$ is defined as a pseudo-differential operator with the symbol
$-|\kappa|^\alpha$ (\cite{Sai1}):
\begin{equation}
\label{spfr}
({\cal F} {d^\alpha\over d|x|^\alpha}f)(\kappa) := -|\kappa|^\alpha \hat 
f(\kappa),
\end{equation}
${\cal F}$ being the Fourier transform of a function
$f$.  For $0<\alpha < 2$, $\alpha\not = 1$, (\ref{spfr}) can be written in the form (\cite{Samko}) 
\begin{equation}
\label{Riesz}
{d^\alpha\over d|x|^\alpha}f(x) ={-1\over 2\Gamma(-\alpha)
\cos(\alpha\pi)}\int_0^\infty {f(x+\xi)-2f(\xi)+f(x-\xi)\over
\xi^{\alpha +1}}d\xi.
\end{equation}
For $\alpha =1$, the relation (\ref{spfr}) can be interpreted in terms of the Hilbert transform 
$$
{d^1\over d|x|^1} f(x) = -{1\over \pi} {d\over dx} \int_{-\infty}^{+\infty}
{f(\xi)\over x-\xi} d\xi,
$$
where the integral is understood in the sense of the Cauchy principal value.
In particular, the equation (\ref{eq}) with $\alpha = 1$ that we call  modified advection equation is written in the form
\begin{equation}
\label{eq-1}
\frac{\partial u}{\partial t}  \, = \,
-{1\over \pi} {d\over dx} \int_{-\infty}^{+\infty}
{f(\xi)\over x-\xi} d\xi
\end{equation}
that is of course different from the standard advection equation.

For $\alpha = 2$, equation (\ref{eq}) is reduced to the one-dimensional wave equation. In what follows, we focus on the case $1\le \alpha < 2$ because the case $\alpha = 2$ (wave equation) is well studied in the literature.

\subsection{Fundamental solution of the fractional wave equation}

We start our analysis by applying the Fourier transform with respect to the space variable $x$ 
to the equation (\ref{eq}) with $1<\alpha <2$ and to the initial conditions (\ref{ic}) with $\varphi (x) = \delta(x)$. Using definition of the Riesz fractional derivative, for the Fourier transform $\hat G_{\alpha}$ we get the initial-value problem
\begin{equation}
\label{4.4}
\left\{
\begin{array}{l}
\hat G(\kappa,0)=1,    \\
\frac{\partial \hat G}{\partial t}(\kappa,0)=0
\end{array} \right.
\end{equation}
for the fractional differential equation 
\begin{equation}
\label{4.3}
(D^{\alpha} \hat G_{\alpha})(t) +|\kappa|^\alpha\hat G_{\alpha}(\kappa,t) =0.
\end{equation}
The unique solution of
(\ref{4.4}), (\ref{4.3}) is given by the expression (see e.g. \cite{Luc99A})
\begin{equation}
\label{4.5}
\hat G_{\alpha}(\kappa,t)  = E_\alpha(-|\kappa|^\alpha t^\alpha)
\end{equation}
in terms of the Mittag-Leffler function 
\begin{equation}
\label{M-L}
E_\alpha(z) =\sum_{k=0}^\infty {z^k\over \Gamma(1+\alpha k)},\ \alpha>0.
\end{equation}

The well known formula (see e.g. \cite{Pod99})
$$
E_\alpha(-x) = -\sum_{k=1}^{m} {(-x)^{-k}\over\Gamma(1-\alpha k)} \ 
+O(|x|^{-1-m}),\ m\in \N,\ x\to +\infty
$$
for asymptotics of the Mittag-Leffler function  that is valid for $0<\alpha <2$ and the formula (\ref{4.5}) show that $\hat G_{\alpha}$ belongs to $L_1(\R)$ with respect to $\kappa$ for $1<\alpha <2$. Therefore we can apply the inverse Fourier transform and get the representation
\begin{equation}
\label{fGR}
G_{\alpha}(x,t)={1\over 2\pi} \int_{-\infty}^{+\infty} 
e^{-i\kappa x}\, E_{\beta}
(-|\kappa|^\alpha t^\beta)\, d\kappa,\ x\in \R,\ t>0
\end{equation}
for the Green function $G_{\alpha}$. 
The last formula shows that the fundamental solution $G_{\alpha}$ is an even function in $x$
$$
G_{\alpha}(-x,t) = G_{\alpha}(x,t),\ x\in \R,\ t>0
$$
and (\ref{fGR})  can be rewritten as the $\cos$-Fourier transform:
\begin{equation}
\label{cosGR}
G_{\alpha}(x,t)={1\over \pi} \int_{0}^{\infty} \cos(\kappa x)
\,E_{\beta}
(-\kappa^\alpha t^\beta)\, d\kappa,\ x\in \R,\ t>0.
\end{equation}
Remarkably, the fundamental solution $G_{\alpha}$ can be represented in terms of elementary functions for every $\alpha,\ 1<\alpha <2$. Indeed, 
for $x=0$ the integral in the right-hand side of (\ref{cosGR}) is reduced to the Mellin 
integral transform of the Mittag-Leffler function at the point 
$s={1\over \alpha}$ (for definition and properties of the Mellin integral transform see e.g. \cite{Mar}). It converges under the conditions  $\alpha >1$ and
its
value 
is given by the formula (\cite{Mar})
\begin{equation}
\label{x=0}
{1\over \pi} \int_{0}^{\infty}
\,E_{\alpha}
(-\kappa^\alpha t^\alpha)\, d\kappa \ =\ {1\over \pi \alpha t}
\int_{0}^\infty E_\alpha(-u)u^{{1\over \alpha}-1}du
\end{equation}
$$
=\ {1\over \pi \alpha t}{\Gamma({1\over \alpha})
\Gamma(1-{1\over \alpha})\over \Gamma(1-\alpha{1\over \alpha})}
\ = \ 0,\ t>0
$$
because the gamma function has a pole at the point $z=0$: $1/\Gamma(0) = 0$. 

Since $G_\alpha$ is an even function, we consider the integral in the right-hand side of (\ref{cosGR}) just in the case $x=|x|>0$ and recognize that this integral is the Mellin convolution of
the functions $g(\xi)=E_\beta(-\xi^\alpha t^\alpha)$ and $f(\xi)
={1\over \pi x \xi}\cos(1/\xi)$ in point $1/x$.
Using the known Mellin integral transforms of the $\cos$-function and the Mittag-Leffler function as well as some elementary properties of the Mellin integral transform (\cite{Mar}) we get the formulas:
$$
g^*(s)=\int_0^\infty g(\tau)\tau^{s-1}d\tau \ = \ 
{1\over \alpha t^{s}}{\Gamma\left({s\over \alpha}\right)
\Gamma\left(1-{s\over \alpha}\right) \over \Gamma(1-s)},\ 
0<\Re(s)<\alpha,
$$
$$
f^*(s) = {1\over \sqrt{\pi} x 2^s}{\Gamma\left({1\over 2}-{s\over 2}\right)\over
\Gamma\left({s\over 2}\right)},\ 0<\Re(s)<1.
$$
These formulas together with the convolution rule and the inverse Mellin integral transform 
lead to the representation
\begin{equation}
\label{H}
G_{\alpha}(x,t)=
{1\over \sqrt{\pi} \alpha x} 
{1\over 2\pi i} \int_{\gamma-i\infty}^{\gamma+i\infty}
{\Gamma\left({s\over \alpha}\right)
\Gamma\left(1-{s\over \alpha}\right) \over \Gamma(1-s)}
{\Gamma\left({1\over 2}-{s\over 2}\right)\over 2^s
\Gamma\left({s\over 2}\right)} \left({t \over x}\right)^{-s}
ds,\ 0<\gamma<1
\end{equation}
of the fundamental solution $G_\alpha$ in terms of the general Fox H-function (see, for example, 
\cite{Mar}, \cite{Samko}, 
\cite{Sc1}). 
The representation (\ref{H}) can be simplified to the form 
\begin{equation}
\label{H1}
G_{\alpha}(x,t)=
{1\over \alpha x} 
{1\over 2\pi i} \int_{\gamma-i\infty}^{\gamma+i\infty}
{\Gamma({s\over \alpha})
\Gamma(1-{s\over \alpha}) \over 
\Gamma(1-{s\over 2})
\Gamma({s\over 2})} \left({t \over x}\right)^{-s}
ds,\ 0<\gamma<\alpha
\end{equation}
and then to the form
\begin{equation}
\label{H2}
G_{\alpha}(x,t)=
{1\over \alpha x} 
{1\over 2\pi i} \int_{\gamma-i\infty}^{\gamma+i\infty}
\frac{\sin(\pi s/2)}{\sin(\pi s/\alpha)}
 \left({t\over x}\right)^{-s}
ds,\ 0<\gamma<\alpha
\end{equation}
by using the duplication and reflection formulas
for the gamma function. 

A useful representation
\begin{equation}
\label{Ka}
G_{\alpha}(x,t)=
{1\over x} L_\alpha (t/x),\ x=|x|>0,\ t>0
\end{equation}
of the fundamental solution $G_\alpha$ in terms of an auxiliary function $L_\alpha$ that depends on the quotient $t/x$ can be obtained from (\ref{H1}) or (\ref{H2}). Moreover, because (\ref{H2}) is in form of an inverse Mellin transform we deduce from (\ref{H2}) the Mellin transform of the function $L_\alpha$ as follows:
\begin{equation}
\label{Ka_Mellin}
L^*_\alpha(s) = \int_0^{\infty} L_\alpha(\tau)\, \tau^{s-1}\, d\tau = \frac{1}{\alpha}\frac{\sin(\pi s/2)}{\sin(\pi s/\alpha)}.
\end{equation}
It follows from (\ref{H2}) that (\ref{Ka_Mellin}) holds true at least for $0<\Re(s)<\alpha$ but as we see later, in fact  (\ref{Ka_Mellin}) is  valid even for $-\alpha<\Re(s)<\alpha$.

Now let us represent the special case (\ref{H1}) of the H-function in form of some convergent series that can be summated in explicit form in terms of some elementary functions. General theory
of the Mellin-Barnes integrals presented e.g. in 
\cite{Mar} says that the integral in (\ref{H1}) is convergent 
under the condition $0<\alpha<2$. For $0<t<x$, the contour of integration in 
the integral (\ref{H1}) can be transformed
to the loop ${\cal L}_{-\infty}$ starting and ending at infinity
and encircling all poles $s_k=-\alpha k,\ k=0,1,2,\dots$ of the
function $\Gamma(s/\alpha)$. Taking into account the relation
$$
\mbox{res}_{s=-k} \Gamma(s) = \frac{(-1)^k}{k!},\ k=0,1,2,\dots,
$$
the residue theorem provides us with the desired series representation:
\begin{equation}
\label{a>b}
G_{\alpha}(x,t)={1\over \alpha x}
\sum_{k=0}^\infty \frac{\alpha (-1)^k}{k!} {\Gamma(1+k)
\over \Gamma\left(-{\alpha\over 2}k\right)\Gamma\left(1-{\alpha\over 2}k\right)}
\left({t\over x}\right)^{\alpha k}
\end{equation}
that can be transformed to the form
\begin{equation}
\label{a>b_1}
G_{\alpha}(x,t)=-{1\over \pi x}
\sum_{k=1}^\infty \sin(\alpha\pi k/2)
\left(-{t^\alpha\over x^\alpha}\right)^{k}
\end{equation}
by using the reflection formula for the gamma function.

Now we use the summation formula
\begin{equation}
\label{im}
\sum_{k=1}^\infty r^k\sin(k a) =
\Im \left( \sum_{k=1}^\infty r^k e^{ik a}\right)  = \Im \left( {r e^{i a}\over 1 - r e^{i
    a}}\right)
 = 
{r\sin a\over 1 -2r\cos a + r^2}
\end{equation}
that is valid for $a\in \R,\ 
|r|<1$ to summate the series in (\ref{a>b_1}) and obtain the nice representation
\begin{equation}
\label{Gel}
G_{\alpha}(x,t)=
{1\over \pi} {x^{\alpha -1} t^\alpha \sin(\pi \alpha/2)\over
t^{2\alpha}+2x^\alpha t^\alpha \cos(\pi\alpha/2)+x^{2\alpha}}
\end{equation}
for the Green function $G_\alpha$ that is valid for $0<t<x$. 

In the case $0<x<t$ we can
transform the contour of integration in (\ref{H1}) to the loop
${\cal L}_{+\infty}$ encircling all poles $s_k = \alpha(1+k),\ 
k=0,1,2,\dots$  of the function
$\Gamma\left(1-{s\over \alpha}\right)$.
Applying the residue theorem  we arrive at the representation
\begin{equation}
\label{a<b}
G_{\alpha}(x,t) =
{1\over \alpha x}
\sum_{k=0}^\infty \frac{\alpha (-1)^k}{k!} {\Gamma(1+k)
\over \Gamma\left({\alpha\over 2}(k+1)\right)\Gamma\left(1-{\alpha\over 2}(k+1)\right)}
\left({x\over t}\right)^{\alpha (k+1)}
\end{equation}
that can be transformed to the form
\begin{equation}
\label{a<b_1}
G_{\alpha}(x,t)=-{1\over \pi x}
\sum_{k=1}^\infty \sin(\alpha\pi k/2)
\left(-{x^\alpha\over t^\alpha}\right)^{k}
\end{equation}
by using the reflection formula for the gamma function. The formula (\ref{im}) applied to the series from the right-hand side of (\ref{a<b_1}) again leads to the representation (\ref{Gel}), this time for  $0<x<t$. Finally, validity of the formula (\ref{Gel}) for $0<x=t$ follows from the principle of analytic continuation for the Mellin-Barnes integrals. Thus
the fundamental solution $G_\alpha$ for the fractional wave equation is given  by the formula
\begin{equation}
\label{Green}
G_{\alpha}(x,t)=
{1\over \pi} {|x|^{\alpha -1} t^\alpha \sin(\pi \alpha/2)\over
t^{2\alpha}+2|x|^\alpha t^\alpha \cos(\pi\alpha/2)+|x|^{2\alpha}},\ t>0,\ x\in \R
\end{equation}
for $1< \alpha <2$.

\subsection{Fundamental solution as a pdf}

We begin by a remark that the formula (\ref{Green}) is valid for $\alpha = 1$ (modified advection equation (\ref{eq-1})), too, that can be proved by direct calculations. In this case we get the well known Cauchy kernel
\begin{equation}
\label{a=1}
G_{1}(x,t)=
{1\over \pi}{t\over t^2+x^2}
\end{equation}
that is a spatial probability density function evolving in time. 

For $\alpha =2$ (wave equation), the Green function $G_2$ is known to be given by the formula
\begin{equation}
\label{a=2}
G_{2}(x,t) = \frac{1}{2}(\delta(x-t)+\delta(x+t)).
\end{equation}
The representation (\ref{Green}) thus leads to an interesting relation
$$
\lim_{\alpha \to 2-0} {|x|^{\alpha -1} t^\alpha \sin(\pi \alpha/2)\over
t^{2\alpha}+2|x|^\alpha t^\alpha \cos(\pi\alpha/2)+|x|^{2\alpha}} 
= \frac{\pi}{2}(\delta(x-t)+\delta(x+t)),\ t>0,\ x\in \R
$$
for the Dirac $\delta$-function.

For $1 < \alpha <2$, the Green function (\ref{Green}) 
is a spatial probability density function evolving in time, too.  Indeed, the function (\ref{Green})  is evidently non-negative for all $t>0$.  Furthermore, for all $t>0$ and $1 < \alpha <2$ the integral
\begin{equation}
\label{int=1}
\int_{-\infty}^{\infty} G_\alpha(x,t)\, dx \, = \, \frac{2}{\pi \alpha} \int_0^{+\infty} \frac{\sin(\pi \alpha/2)}{
1+2u \cos(\pi\alpha/2)+u^2}\, du 
\end{equation}
is identically equal to $1$ that can be checked by direct calculations. Thus $G_\alpha$ given by (\ref{Green}) is a spatial probability density function evolving in time that can be considered to be a fractional generalization of the Cauchy kernel (\ref{a=1}) for the case of an arbitrary index $\alpha,\ 1\le \alpha < 2$.

Now let us study the properties of the fundamental solution (\ref{Green}) as a pdf. Because $G_\alpha$ is an even function we can restrict our attention to the case $x \ge 0$ and consider the function
$$
G^{+}_{\alpha}(x,t)=
{1\over \pi } {x^{\alpha-1} t^\alpha \sin(\pi \alpha/2)\over
t^{2\alpha}+2x^\alpha t^\alpha \cos(\pi\alpha/2)+x^{2\alpha}},\ x\ge 0,\ t>0,\ 1<\alpha<2.
$$

It is easy to see that  $G_\alpha^+$ behaves like a power function in $x$ both at $x=0$ and at $x=+\infty$ for a fixed $t>0$:
\begin{equation}
\label{asym}
G_{\alpha}^+(x,t) \approx \begin{cases}
x^{\alpha-1},\ x\to 0, \\
x^{-\alpha-1},\ x\to +\infty.
\end{cases}
\end{equation}
This means that the pdf $G_\alpha$ possesses all finite moments up to the order $\alpha$. In particular, the mean value of $G_\alpha$ (its first moment) exists for all $\alpha > 1$ (we note that the Cauchy kernel does not possess a mean value). Let us now evaluate the moments of the one-sided fractional Cauchy kernel $G_\alpha^+$ for a fixed $t>0$. To do this, we refer to the representation (\ref{Ka}) of $G_\alpha^+$ in terms of the auxiliary function $L_\alpha$ that can be now represented in the form
\begin{equation}
\label{La}
L_\alpha (\tau) = \frac{1}{\pi} \frac{\tau^\alpha  \sin(\pi \alpha/2)}
{\tau^{2\alpha}+2\tau^\alpha \cos(\pi\alpha/2)+1},\ \tau>0,\ 1<\alpha<2.
\end{equation}
Taking into account this formula, the function $\frac{1}{|\tau|}L_\alpha(|\tau|)$ can be interpreted as a fractional Cauchy pdf of order $\alpha$. 

The moment of the order $\beta,\ |\beta| <\alpha$ of $G_\alpha^+$ can be represented in terms of the Mellin integral transform of  $L_\alpha$ that is known (see the formula (\ref{Ka_Mellin})) and thus evaluated:
\begin{equation}
\label{mom}
\int_0^\infty G_{\alpha}^+(x,t)x^\beta\, dx = t^\beta \int_0^\infty L_\alpha(\tau)\tau^{-\beta-1}\, d\tau = \frac{t^\beta }{\alpha} \frac{\sin(\pi \beta/2)}
{\sin(\pi\beta/\alpha)}.
\end{equation}
In particular, 
we get the formula
\begin{equation}
\label{0mom}
\int_0^\infty G_{\alpha}^+(x,t)\, dx =\frac{1}{2}
\end{equation}
that is in accordance with (\ref{int=1}) because $G_\alpha$ is an even function. 

We mention also important formula
\begin{equation}
\label{1mom}
\int_0^\infty G_{\alpha}^+(x,t)x\, dx = \frac{t}{\alpha \sin(\pi/\alpha)},\ 1<\alpha\le 2
\end{equation}
for the first moment of the one-sided fractional Cauchy kernel $G_\alpha^+$. 

\subsection{Extrema points, gravity and mass centers of $G_\alpha$, and location of its energy}

Now we derive some important analytical properties of the fractional Cauchy kernel (\ref{Green}). First we remark that $G_\alpha(0,t)=0$ and $G_\alpha(x,t)>0$ for $x\not = 0$, so that  $x=0$ is a minimum point for the fundamental solution $G_\alpha$ for any $t>0$. Because $G_\alpha$ is an even function we again consider its restriction to $x\ge 0$, i.e. the function 
$G^{+}_{\alpha}$. 

To determine locations of maxima of $G^{+}_{\alpha}$ for fixed values of $t$ and $\alpha$ we solve the equation
$$
\frac{\partial G^{+}_{\alpha}}{\partial x}(x,t) = 0
$$
that  turns out to be equivalent to the quadratic equation
$$
(\alpha +1)\left(\frac{x^\alpha}{t^\alpha}\right)^2 + 2\cos(\pi\alpha/2) \left(\frac{x^\alpha}{t^\alpha}\right) - (\alpha -1) = 0
$$
with solutions given by
$$
\frac{x^\alpha}{t^\alpha} = \frac{-\cos(\pi\alpha/2) \pm \sqrt{\alpha^2 - \sin^2(\pi\alpha/2)}}{\alpha +1}.
$$
Since we are interested in nonnegative solutions the only candidate for this role is the point
\begin{equation}
\label{c_a}
\frac{x^\alpha}{t^\alpha} = c_\alpha,\ c_\alpha:=\frac{-\cos(\pi\alpha/2) + \sqrt{\alpha^2 - \sin^2(\pi\alpha/2)}}{\alpha +1}.
\end{equation}
Because $\frac{\partial G^{+}_{\alpha}}{\partial x}(x,t)$ is positive for 
$\frac{x^\alpha}{t^\alpha} < c_\alpha$ and negative for $\frac{x^\alpha}{t^\alpha} > c_\alpha$ we conclude that the point
\begin{equation}
\label{loc_max}
x_\alpha^\star(t) = v_p(\alpha) t,\ v_p(\alpha):=(c_\alpha)^{\frac{1}{\alpha}}
\end{equation}
with $c_\alpha$ given by (\ref{c_a}) is the only maximum point of the fractional Cauchy kernel $G_\alpha$ for  $x\ge 0$. Of course, the point $-x_\alpha^\star(t) < 0$ is another maximum point of $G_\alpha$ because $G_\alpha$ is an even function.

To determine the maximum value of the function $G_\alpha$ that coincides with the maximum value of $G_\alpha^+$ and is denoted by  $G_\alpha^\star(t)$ we substitute the point $x=x_\alpha^\star(t)$ given by (\ref{loc_max}) into the function $G_\alpha^+$ and get
\begin{equation}
\label{max}
G_\alpha^\star(t) = \frac{1}{\pi v_p(\alpha) t}\frac{ c_\alpha \sin(\pi \alpha/2)}
{1+2c_\alpha \cos(\pi\alpha/2)+c_\alpha^2},
\end{equation}
where $v_p(\alpha)$ and $c_\alpha$ are defined as in the formulas (\ref{c_a}) and (\ref{loc_max}). 

It follows from the formulas (\ref{loc_max}) and (\ref{max}) that for a fixed value of $\alpha, 1<\alpha < 2$ the product $p_\alpha$ of the maximum value $G_\alpha^\star(t)$ and the maximum location $x_\alpha^\star(t)$ is time-independent:
\begin{equation}
\label{prod}
p_\alpha = G_\alpha^\star(t)\cdot x_\alpha^\star(t) = \frac{1}{\pi } \frac{ c_\alpha \sin(\pi \alpha/2)}
{1+2c_\alpha \cos(\pi\alpha/2)+c_\alpha^2}.
\end{equation}
This means that the maximum point $(x_\alpha^\star(t),\, G_\alpha^\star(t))$ of the fundamental solution $G_\alpha$ with a fixed value of $\alpha, 1<\alpha < 2$ moves in time along a hyperbola that is completely determined by the value of $\alpha$ (see the formula (\ref{prod})). It is interesting to note that the product $p_\alpha$ coincides with the value of the fundamental solution $G_\alpha$ at the point $(1,v_p(\alpha))$:
$$
p_\alpha = \frac{1}{\pi } \frac{ c_\alpha \sin(\pi \alpha/2)}
{1+2c_\alpha \cos(\pi\alpha/2)+c_\alpha^2} = G_\alpha(1,v_p(\alpha)).
$$
This property can be probably used to give a physical interpretation of the formula (\ref{prod}).

Now we calculate the location of the gravity center $x_\alpha^{g}(t)$ of the fundamental solution $G_\alpha$ that is defined  by the formula (we recall that $G_\alpha$ is an even function) 
\begin{equation}
\label{grav}
x_\alpha^{g}(t) = \frac{\int_0^\infty x\, G_\alpha(x,t)\, dx}
{\int_0^\infty G_\alpha(x,t)\, dx}.
\end{equation}
Using the formulas (\ref{0mom}) and (\ref{1mom})  we get the following result:
\begin{equation}
\label{grav_f_x}
x_\alpha^{g}(t) = 
\frac{2t}{\alpha \sin(\pi/\alpha)}.
\end{equation}
The "mass"-center $x_\alpha^{m}(t)$ of $G_\alpha$ is determined by the formula (\cite{Gur01})
\begin{equation}
\label{mass}
x_\alpha^{m}(t) = \frac{\int_0^\infty x\, G_\alpha^2(x,t)\, dx}
{\int_0^\infty G_\alpha^2(x,t)\, dx}.
\end{equation}
Substituting (\ref{Green}) into (\ref{mass}) and transforming the obtained integrals we get the representation
\begin{equation}
\label{mass_f_x}
x_\alpha^{m}(t) = v_m(\alpha)\, t,\ \ v_m(\alpha) = 
\frac{\int_0^\infty \tau^{-1}\, L_\alpha^2(\tau)\, d\tau}
{\int_0^\infty \tau^{-2}\, L_\alpha^2(\tau)\, d\tau},
\end{equation}
where the function $L_\alpha$ is defined by (\ref{La}). Because the Mellin transform of $L_\alpha$ is known (see the formula (\ref{Ka_Mellin})), the integrals 
\begin{equation}
\label{intL}
{\int_0^\infty \tau^{\beta}\, L_\alpha^2(\tau)\, d\tau},\ -2\alpha -1<\beta <2\alpha -1
\end{equation}
can be interpreted as Mellin convolutions of the functions $\tau^{\beta+1}L_\alpha(\tau)$ and $L_\alpha(1/\tau)$ in point $x=1$ and thus expressed in terms of an H-function with parameters depending on $\alpha$ and $\beta$ and with the argument $x=1$. Because there are no routines for numerical evaluation of the H-function available, we prefer to stay by the representation of $v_m(\alpha)$ given in (\ref{mass_f_x}) and not to transform it to a quotient of two H-functions. Remarkably, there exists an explicit formula for the integrals (\ref{intL}) in the case $\alpha = 1$ (modified advection equation), namely (\cite{Pru86})
\begin{equation}
\label{intL1}
{\int_0^\infty \tau^{\beta}\, L_1^2(\tau)\, d\tau}=\frac{1+\beta}{4\pi}\frac{1}{\cos(\pi\beta/2)},\ -3<\beta <1.
\end{equation}
The "mass"-center $x_1^{m}$ of $G_1$ can be then represented by the simple formula
\begin{equation}
\label{mass_f_x_1}
x_1^{m}(t) = \frac{2}{\pi}\, t.
\end{equation}

Finally we mention that "location of energy" of the damped wave that is represented by the fundamental solution $G_\alpha$ is given by the formula (\cite{Car10})
\begin{equation}
\label{energy}
t_\alpha^{c}(x) = \frac{\int_0^\infty t\, G_\alpha^2(x,t)\, dt}
{\int_0^\infty G_\alpha^2(x,t)\, dt}
\end{equation}
and can be represented in the form
\begin{equation}
\label{energy_f_x}
t_\alpha^{c}(x) = \frac{x}{v_c(\alpha)},\ \ v_c(\alpha) = 
\frac{\int_0^\infty  L_\alpha^2(\tau)\, d\tau}
{\int_0^\infty \tau\, L_\alpha^2(\tau)\, d\tau},
\end{equation}
where the function $L_\alpha$ is defined by (\ref{La}). Because the integral $\int_0^\infty  \tau\, L_\alpha^2(\tau)\, d\tau$ diverges for $\alpha = 1$, $v_c(\alpha)$ tends to $0$ as $\alpha \to 1$. 
 
\subsection{The velocities of the damped waves}

It is well known (see e.g. \cite{Blo77}, \cite{Car10}, \cite{Gur01}, or \cite{Smi70}) that several different definitions of the wave velocities and in particular of light velocity  can be introduced. For the damped waves that are described by the fractional wave equation (\ref{eq}) we calculate propagation velocity of the maximum of its fundamental solution $G_\alpha$ that can be interpreted as the phase velocity, propagation velocity of the gravity center of $G_\alpha$, the velocity of its "mass"-center or the pulse velocity, and three different kinds of its centrovelocity. It turns out that all these velocities are constant in time and depend just on the order $\alpha$ of the fractional wave equation. Whereas four out of six velocities are different each to other, the fist centrovelocity coincides with the Smith centrovelocity and the the second centrovelocity is the same as the pulse velocity. 

We start with the phase velocity and determine it using the formula (\ref{loc_max}) that leads to the result that the maximum location of the fundamental solution $G_\alpha$ propagates with a constant velocity $v_p(\alpha)$ that is given by the expression
\begin{equation}
\label{speed}
v_p(\alpha) := \frac{d x_\alpha^\star(t)}{dt}  = \left(\frac{-\cos(\pi\alpha/2) + \sqrt{\alpha^2 - \sin^2(\pi\alpha/2)}}{\alpha +1}\right)^{\frac{1}{\alpha}}.
\end{equation}
For $\alpha = 1$ (modified advection equation (\ref{eq-1})), the propagation velocity of the maximum of $G_\alpha$ is equal to zero (the maximum point stays at $x=0$) whereas for $\alpha = 2$ (wave equation) the maximum point propagates with the constant velocity $1$. 

To determine the propagation velocity $v_g(\alpha)$ of the gravity center of $G_\alpha$ we employ  the formula (\ref{grav_f_x}) and get the following result:
\begin{equation}
\label{grav_f}
v_g(\alpha) := \frac{d x_\alpha^{g}(t)}{dt} = 
\frac{2}{\alpha \sin(\pi/\alpha)}.
\end{equation}
$v_g(\alpha)$ is thus time-independent and determined by the order $\alpha$ of the fractional wave equation. Evidently, $v_g(2)=1$ and $v_g(\alpha)\to +\infty$ as $\alpha \to 1+0$.   

The velocity $v_m(\alpha)$ of the "mass"-center of $G_\alpha$ or its pulse velocity (\cite{Gur01}) is obtained from the formula (\ref{mass_f_x}) and is equal to 
\begin{equation}
\label{mass_f}
v_m(\alpha):= \frac{dx_\alpha^{m}(t)}{dt} = 
\frac{\int_0^\infty \tau^{-1}\, L_\alpha^2(\tau)\, d\tau}
{\int_0^\infty \tau^{-2}\, L_\alpha^2(\tau)\, d\tau},
\end{equation}
where the function $L_\alpha$ is defined by (\ref{La}). For $\alpha =1$, the pulse velocity is equal to $\frac{2}{\pi} \approx 0.64$ (see the formula (\ref{mass_f_x_1})). 

Following \cite{Car10} we define the second centrovelocity $v_2(\alpha)$ as the mean pulse velocity computed from $0$ to time $t$. It follows from (\ref{mass_f_x}) and (\ref{mass_f}) that for the damped wave that is described by the fundamental solution of the fractional wave equation the second centrovelocity is equal to its pulse velocity $v_m(\alpha)$:
\begin{equation}
\label{centro_2}
v_2(\alpha):= \frac{x_\alpha^{m}(t)}{t} = v_m(\alpha) = 
\frac{\int_0^\infty \tau^{-1}\, L_\alpha^2(\tau)\, d\tau}
{\int_0^\infty \tau^{-2}\, L_\alpha^2(\tau)\, d\tau}.
\end{equation}

The Smith centrovelocity $v_c(\alpha)$ (\cite{Smi70}) of the damped waves describes the motion of the first moment of their energy distribution and can be evaluated in explicit form using the formula (\ref{energy_f_x}):
\begin{equation}
\label{energy_f}
v_c(\alpha):= \left(\frac{dt_\alpha^{c}(x)}{dx}\right)^{-1} = 
\frac{\int_0^\infty  L_\alpha^2(\tau)\, d\tau}
{\int_0^\infty \tau\, L_\alpha^2(\tau)\, d\tau},
\end{equation}
where the function $L_\alpha$ is defined by (\ref{La}). Because the integral $\int_0^\infty  \tau\, L_\alpha^2(\tau)\, d\tau$ diverges for $\alpha = 1$, the Smith centrovelocity tends to $0$ as $\alpha \to 1$.

Finally, we calculate the first centrovelocity $v_1(\alpha)$ that is defined as the mean centrovelocity from $0$ to $x$ (\cite{Car10}). It follows from (\ref{energy_f_x}) and (\ref{energy_f}) that for the damped wave $G_\alpha$  the first centrovelocity is equal to the Smith centrovelocity $v_c(\alpha)$:
\begin{equation}
\label{centro_1}
v_1(\alpha):= \frac{x}{t_\alpha^{c}(x)} = v_c(\alpha) = 
\frac{\int_0^\infty  L_\alpha^2(\tau)\, d\tau}
{\int_0^\infty \tau\, L_\alpha^2(\tau)\, d\tau}.
\end{equation}

As we have seen, all velocities introduced above are constant in time and depend just on the order $\alpha$ of the fractional wave equation. The phase velocity, the velocity of the gravity center of $G_\alpha$, the pulse velocity, and the Smith centrovelocity are different each to other whereas the fist centrovelocity coincides with the Smith centrovelocity and the second centrovelocity is the same as the pulse velocity. For the physical interpretation and meaning of the velocities that were determined above we refer to e.g. \cite{Blo77}, \cite{Car10}, or \cite{Gur01}.

\section{Discussion of the obtained results and plots}

\begin{figure}
\begin{center}
\includegraphics[width=6cm, height=3cm]{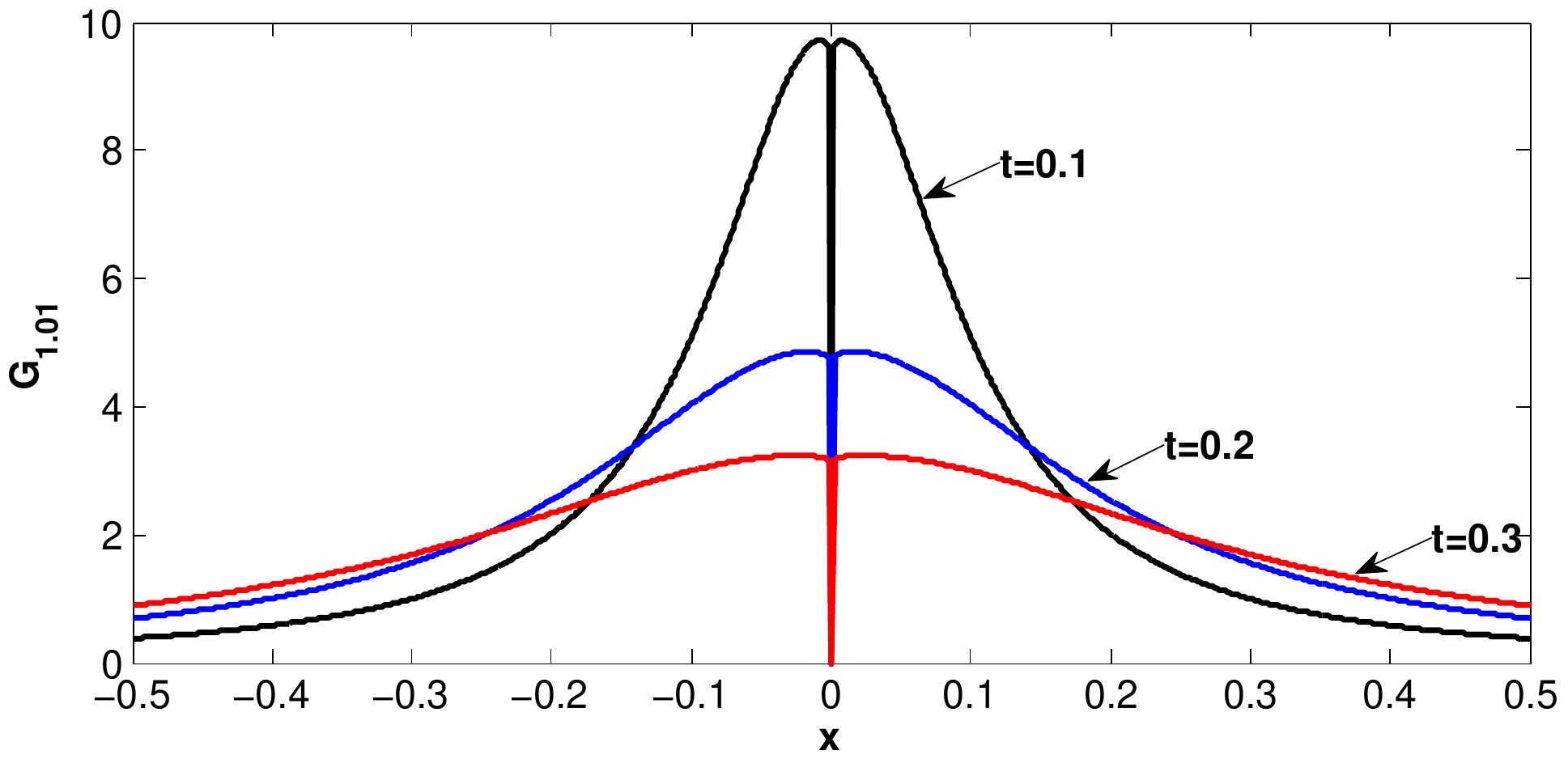}   
\includegraphics[width=6cm, height=3cm]{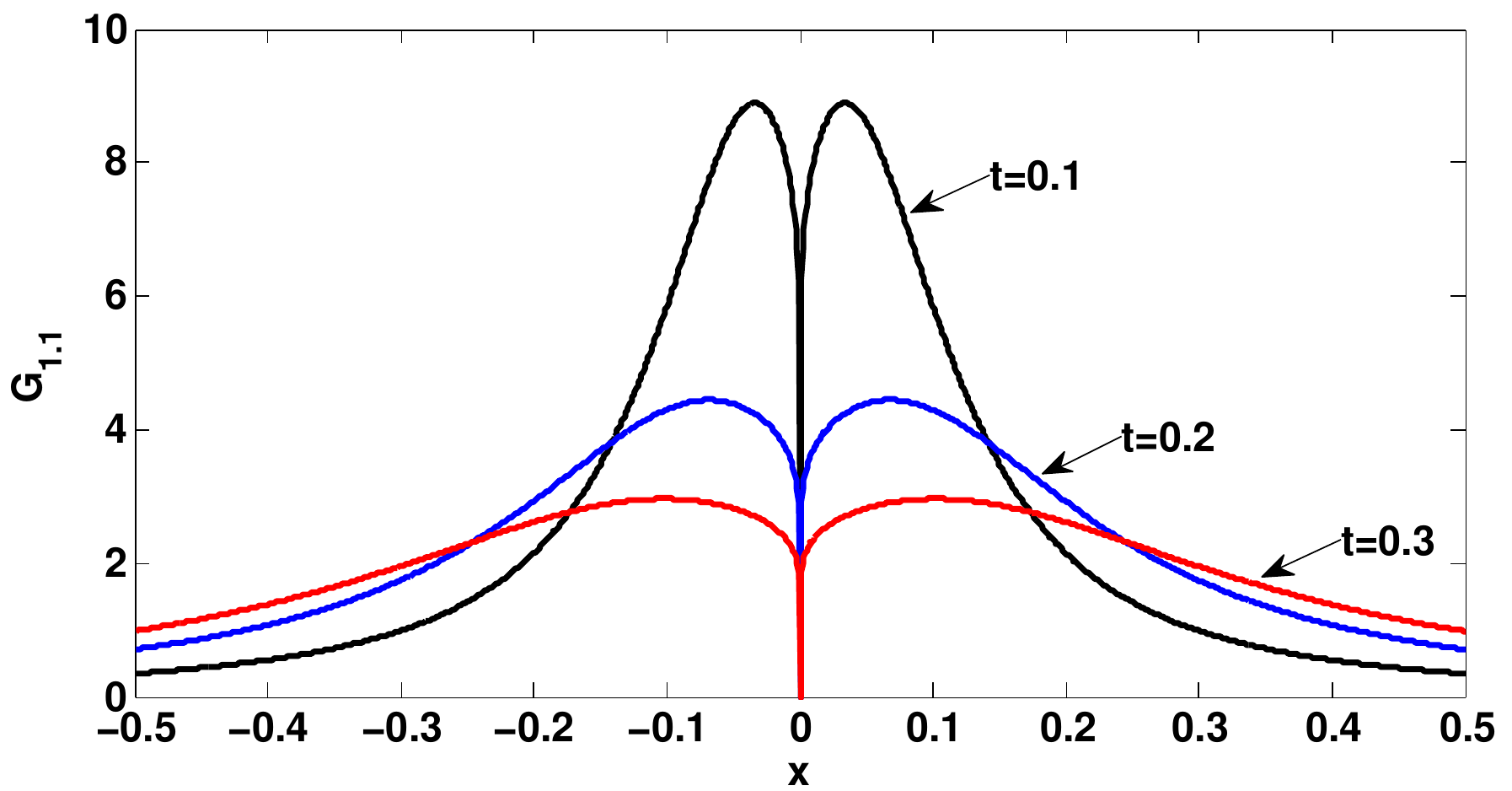}   
\includegraphics[width=6cm, height=3cm]{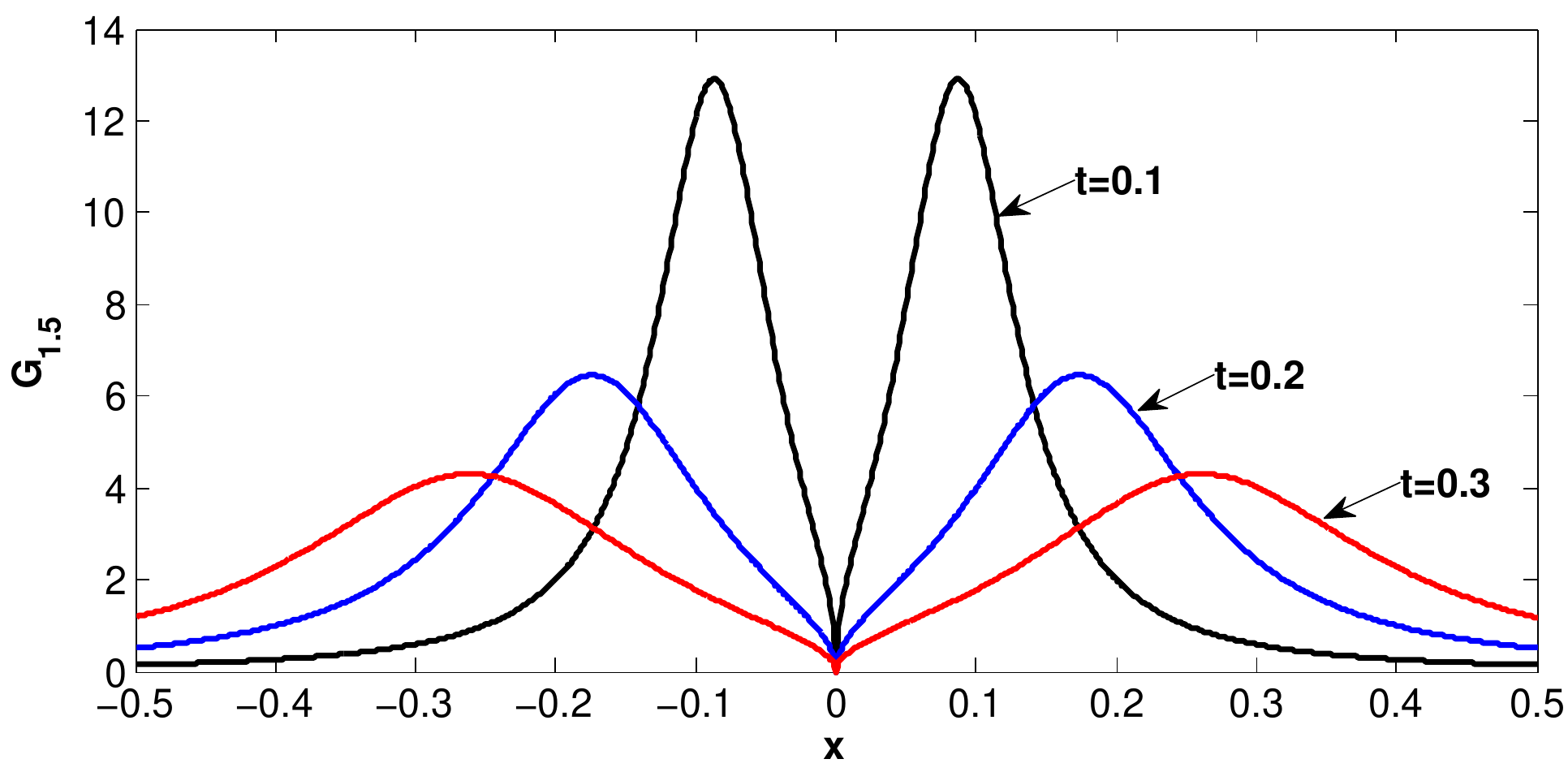}   
\includegraphics[width=6cm, height=3cm]{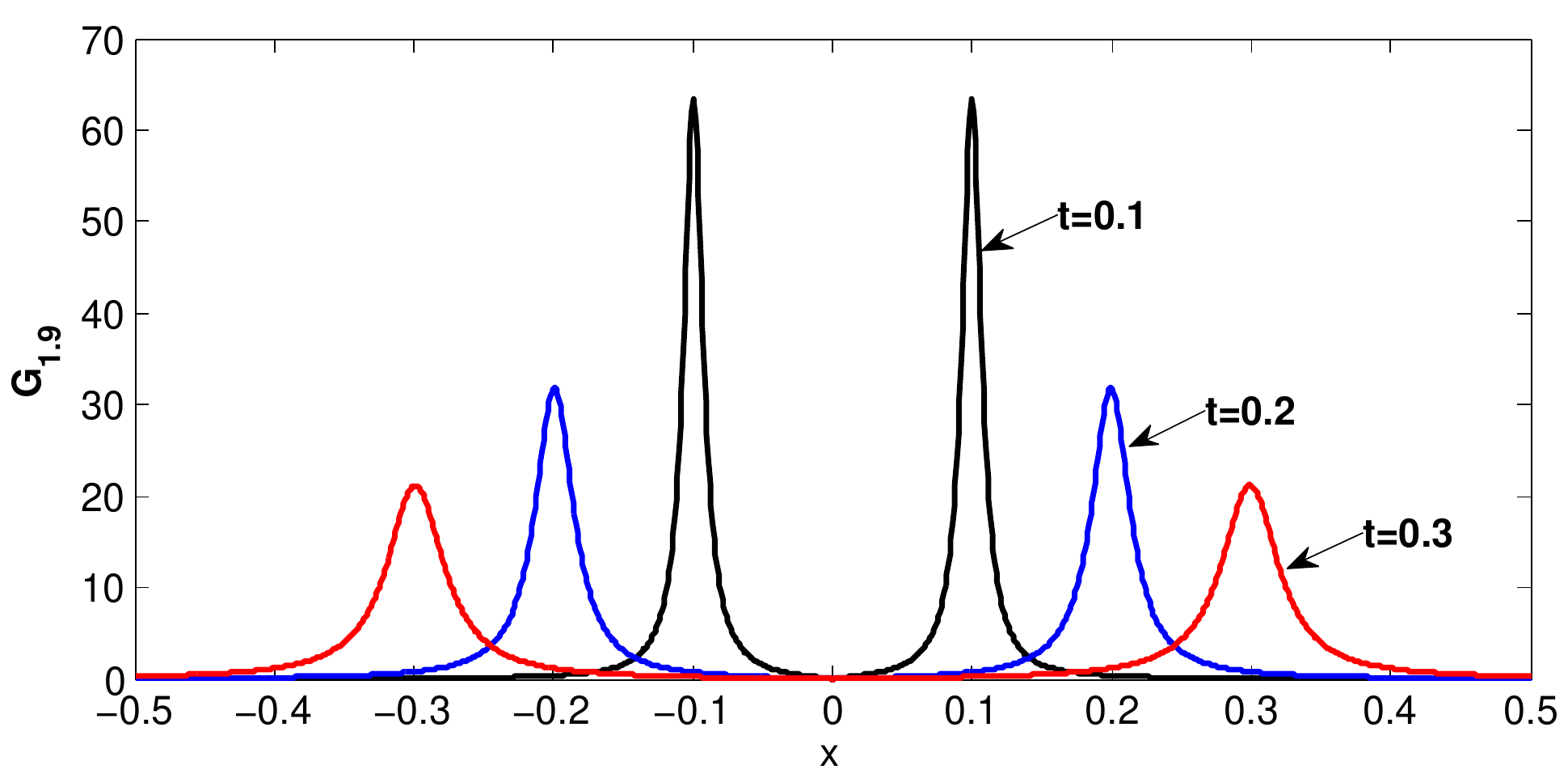}  
\caption{Fundamental solution $G_\alpha$: Plots for values of $\alpha = 1.01$ (1st line, left), $1.1$ (1st line, right), $1.5$ (2nd line left), and $1.9$ (2nd line, right) for $-0.5\le x\le 0.5$ and $t=0.1,\, 0.2,\ 0.3$ }
\label{fig-1}
\end{center}
\end{figure}

\begin{figure}
\begin{center}
\includegraphics[width=8cm]{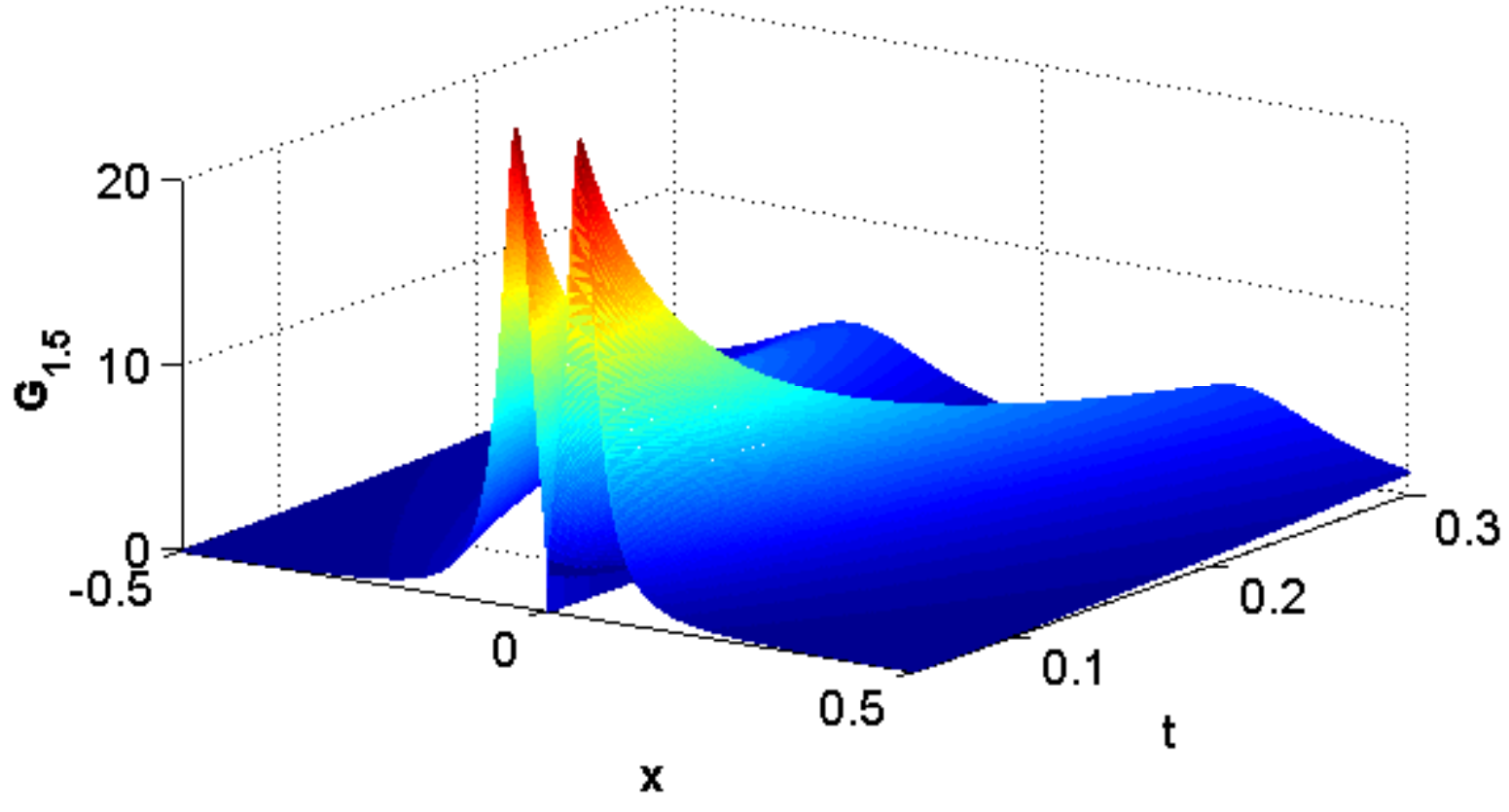}    
\includegraphics[width=4cm]{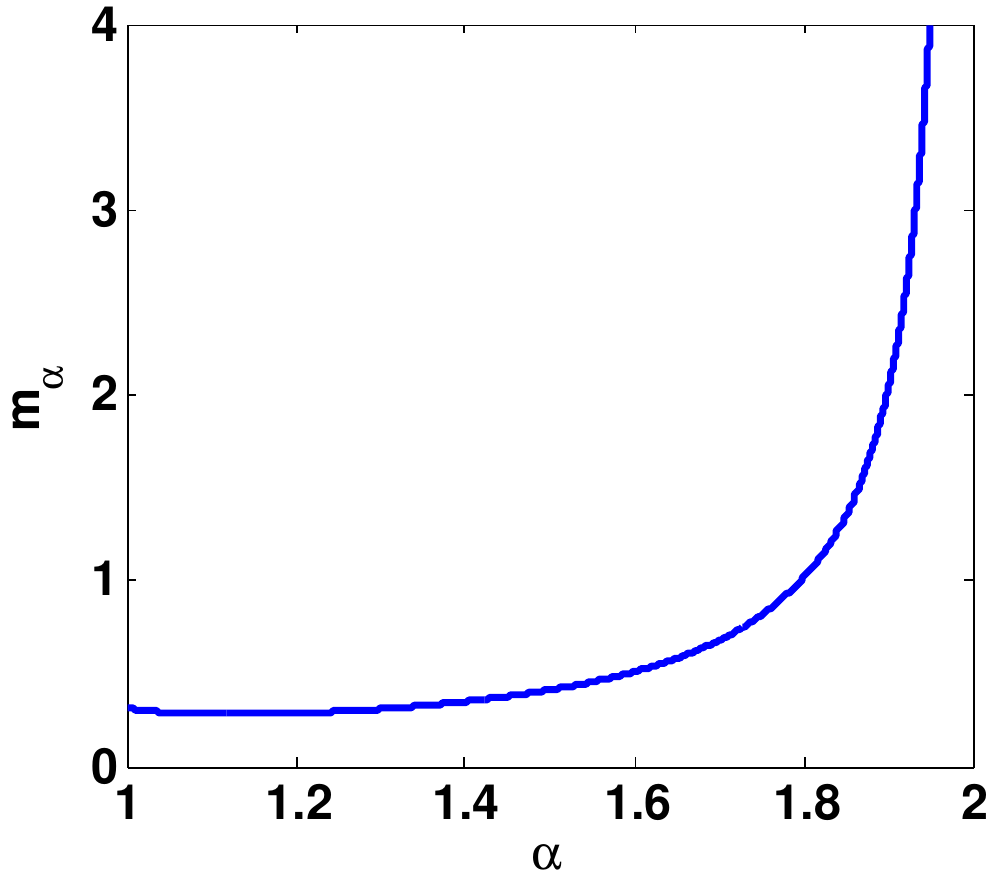}   
\caption{Plots of $G_\alpha$ for $\alpha = 1.5$, $-0.5\le x\le 0.5$, and $0<t\le 0.3$ (left) and of the maximum value $m_\alpha$ of the function $G_\alpha$ at the time instant $t=1$ (right) }
\label{fig-3}
\end{center}
\end{figure}

To start with, let us consider evolution of the fundamental solution $G_\alpha$ in time  for some characteristic values of $\alpha$. In Fig. \ref{fig-1} plots of $G_\alpha$ for $\alpha = 1.01,\, 1.1,\, 1.5$, and $1.9$ are presented. As we can see, in all cases maximum location is moved linearly in time according to the formula (\ref{loc_max}) whereas the maximum value decreases according to the formula (\ref{max}). The behavior of $G_\alpha$ can be thus interpreted as propagation of damped waves whose amplitude decreases with time. 
This phenomena can be very clearly recognized  on the 3D plot presented in Fig. \ref{fig-3}. Of course, because of the nonlocal character of the fractional derivatives in the fractional wave equation solutions to this equation show some properties of diffusion processes, too. In particular, the fundamental solution $G_\alpha$ is positive for all $x\not = 0$ at any small time instance $t>0$ that means that a disturbance of the initial conditions spreads infinitely fast and the equation (\ref{eq}) is non relativistic like the classical diffusion equation. But in contrast to the diffusion equation, both the maximum of the fundamental solution $G_\alpha$, its gravity and mass centers and location of its energy propagate with the finite constant velocities like in the case of the fundamental solution of the wave equation. 

\begin{figure}
\begin{center}
\includegraphics[width=6cm]{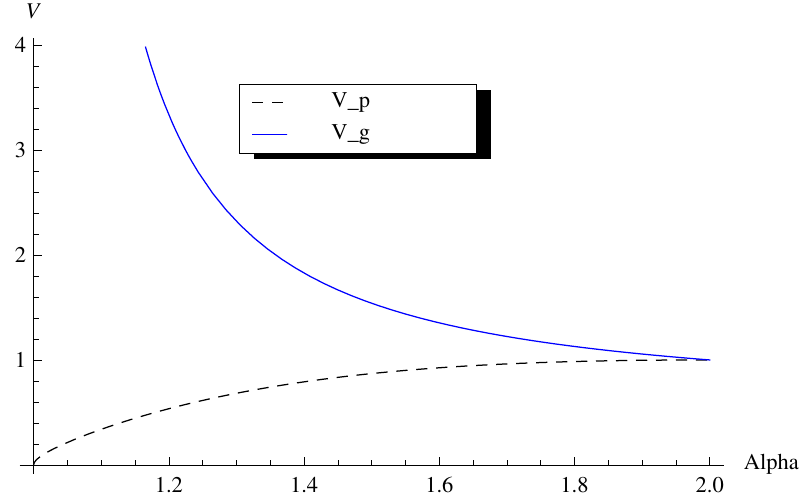}   
\includegraphics[width=6cm]{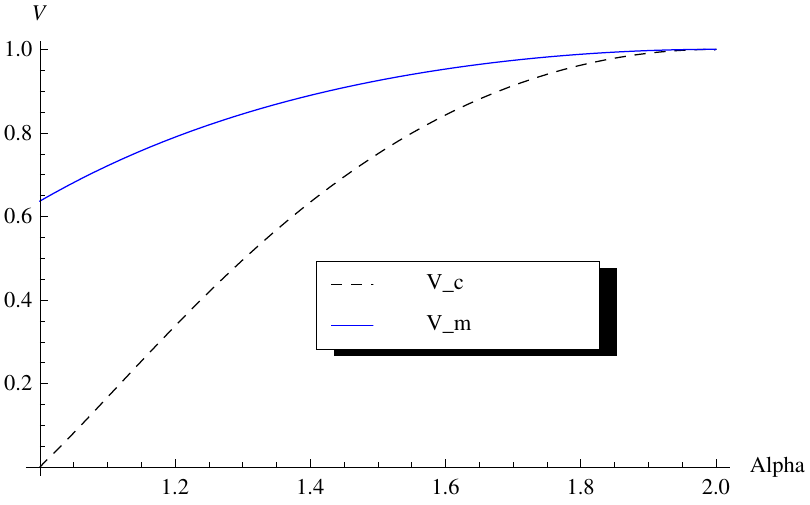}    
\caption{Plots of the phase velocity $v_p(\alpha)$ and the gravity center velocity $v_g(\alpha)$ (left) and of the pulse velocity $v_m(\alpha)$ and the centrovelocity $v_c(\alpha)$ (right)}
\label{fig-4}
\end{center}
\end{figure}

Plots of the propagation velocity $v_p$ of the maximum of the fundamental solution $G_\alpha$ (phase velocity), the velocity $v_g$ of its gravity center, its pulse velocity $v_m$ and its centrovelocity $v_c$ are presented in Fig. \ref{fig-4}. As expected, $v_p=v_c=0$ and $v_m\approx 0.64$ for $\alpha = 1$ (modified advection equation) and all velocities smoothly approach the value $1$ as $\alpha \to 2$ (wave equation). For $1<\alpha <2$, $v_p,\ v_m,$ and $v_c$ monotonously  increase whereas $v_g$ monotonously  decreases. It is interesting to note that for all velocities the property $\frac{dv(\alpha)}{d\alpha}(2-0) = 0$ holds true, i.e. in a small neighborhood of the point $\alpha =2$ the velocities of $G_\alpha$ are nearly the same as in the case of the fundamental solution of the wave equation. The velocity $v_g$ of the gravity center of $G_\alpha$ tends to $+\infty$ for $\alpha \to 1+0$ and $t>0$ (modified advection equation) because the first moment of the Cauchy kernel (\ref{a=1}) does not exist. It is interesting to note that for all $\alpha, 1<\alpha <2$ the velocities $v_p,\ v_g,\ v_m,\ v_c$ are different each to other and fulfill the inequalities $v_c(\alpha) <v_p(\alpha) <v_m(\alpha) <v_g(\alpha)$. For $\alpha = 2$, all velocities are equal to 1. 

In Fig. \ref{fig-3} (right), a plot of the maximum value of the fundamental solution $G_\alpha$ that is given by the formula (\ref{max}) is presented for $t=1$.  Surprisingly, the function $m_\alpha = G_\alpha^\star(1)$ is not monotone. Numerical calculations show that it has a minimum that is located at the point $\alpha \approx 1.13$. At $\alpha = 1$, $G_\alpha(x,1)$ has a  (local) maximum with the value $\frac{1}{\pi}\approx 0.32$. Then $G_\alpha(x,1)$ monotonically decreases to $m_{\mbox{{\small min}}} \approx 0.28$ at $\alpha_{\mbox{{\small min}}} \approx 1.13$ (minimum) and then starts to monotonically increase and tends to $+\infty$ when $\alpha \to 2-0$ that is in accordance with behavior of the fundamental solution $(\frac{1}{2}(\delta(x-t)+\delta(x+t))$ of the wave equation ($\alpha =2$). 
We note that the location $x_\alpha^*(t)$ of the maximum value $G_\alpha^*(t)$ of $G_\alpha$ is given by the formula (\ref{loc_max}). For  $t=1$,  location of the maximum value coincides with the phase velocity $v_p$ that is presented in Fig. \ref{fig-4}. 
\begin{figure}
\begin{center}
\includegraphics[width=5cm, height=5cm]{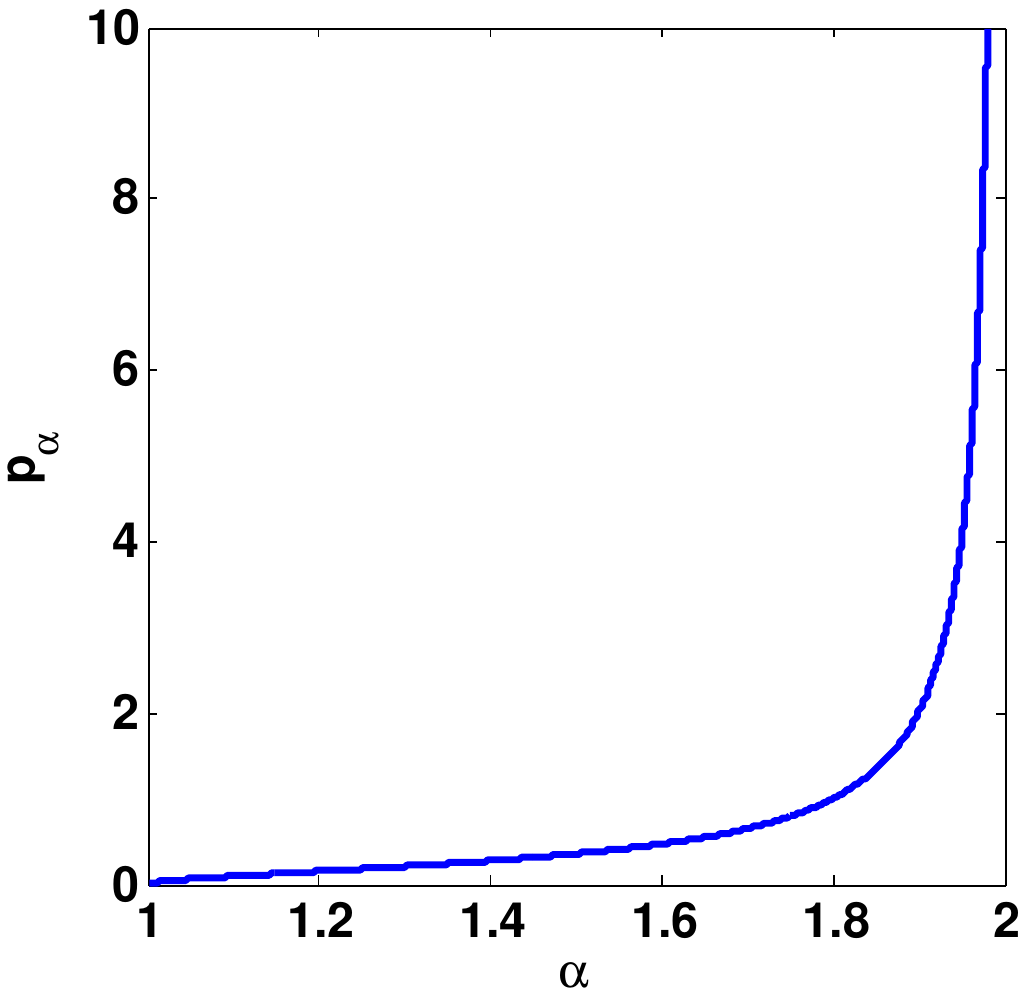}   
\includegraphics[width=5cm, height=5cm]{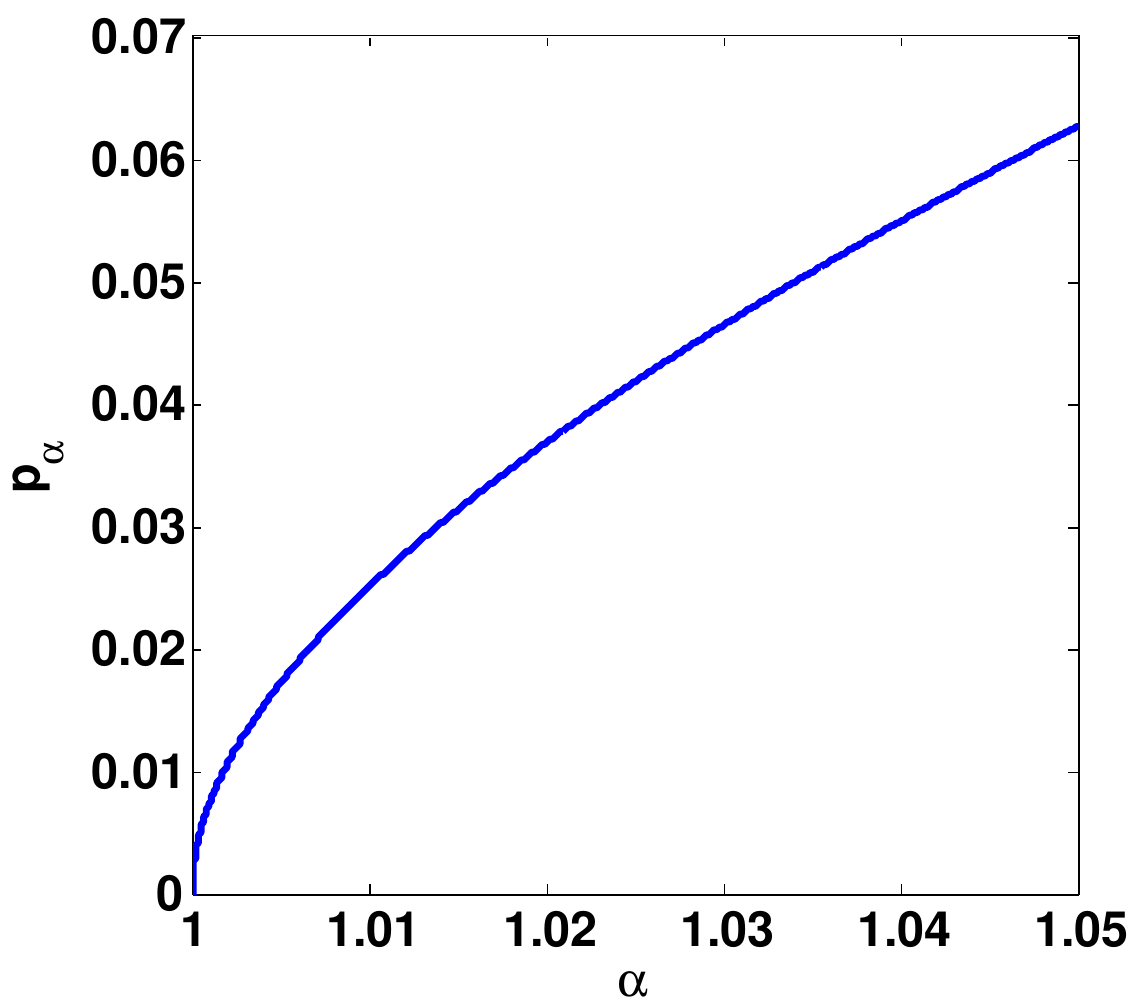}    
\caption{Plots of the product $p_\alpha$ of the maximum value $G_\alpha^\star(t)$ and the maximum location $x_\alpha^\star(t)$ on the interval $1\le \alpha < 2$ (left) and on the interval $1\le \alpha \le 1.05$ (right)}
\label{fig-5}
\end{center}
\end{figure}

Finally, some plots of the product $p_\alpha$ of the maximum value $G_\alpha^\star(t)$ and the maximum location $x_\alpha^\star(t)$ given by the formula (\ref{prod}) are presented in Fig. \ref{fig-5}. The function $p_\alpha = p_\alpha(\alpha)$ is monotone increasing for $1\le \alpha < 2$ and varies from $0$ at the point $\alpha = 1$ (the location of the maximum of the Cauchy kernel (\ref{a=1}) is always at the point $x=0$) to $+\infty$ at the point $\alpha = 2$ (the maximum value of the the Green function $(\frac{1}{2}(\delta(x-t)+\delta(x+t))$ of the wave equation is $+\infty$). In this sense, the product $p_\alpha$ can be considered to be a  characteristic property of the damped waves that are described by the fractional wave equation (\ref{eq}). As we can see on the left plot, the product $p_\alpha$ varies between 0 and 10 for $1\le \alpha \le 1.98$ and changes very slowly for $1.01 \le \alpha \le 1.9$. For $\alpha \to 1+0$ (see the right plot of Fig. \ref{fig-5}) and $\alpha \to 2-0$, $p_\alpha$ goes to 0 and $+\infty$, respectively, very fast. 
For a damped wave that is supposed to be described by the fractional wave equation (\ref{eq}) with an unknown exponent $\alpha$, the value of the product $p_\alpha$ can be measured and used to recover $\alpha$ either from the formula (\ref{prod}) or graphically from Fig. \ref{fig-5}. 
 
\section{Conclusions and open questions}
In this paper, a model one-dimensional fractional wave equation with the fractional derivatives of order $\alpha,\ 1\le \alpha \le 2$ both in space and in time is introduced and considered in detail.  The fractional wave equation inherits some crucial characteristics of the wave equation like  constant propagation velocities of the maximum of its Green function, its gravity and mass centers, and its energy location. Because the maximum value of the fundamental solution $G_\alpha$ (wave amplitude) decreases with time whereas its location moves with a constant velocity, solutions to the fractional wave equation can be interpreted as damped waves.  Moreover, $G_\alpha$ that turns out to be expressed in terms of elementary functions for all values of $\alpha,\ 1\le \alpha < 2$ can be interpreted as a spatial probability density function evolving in time whose moments up to order $\alpha$ are finite. For the fundamental solution $G_\alpha$, both its maximum location and its maximum value are determined in closed form.  Remarkably, the product of the maximum location and the maximum value of $G_\alpha$  is time-independent and just a function of $\alpha$. 

Among problems for further research we mention two- and three-dimensional fractional wave equations with different initial or/and boundary conditions. Of course, it would be interesting to consider fractional wave equations with fractional derivatives defined in different ways. We note here that in \cite{MaiLucPag} a space-time fractional diffusion-wave equation with the Caputo derivative of order $\beta \in (0,2]$ in time and the Riesz-Feller derivative of order $\alpha \in (0,2]$ and skewness $\theta$ in space has been investigated in detail. A particular case of this equation called neutral-fractional diffusion equation that for $\theta=0$ corresponds to our fractional diffusion equation  has been shortly mentioned in \cite{MaiLucPag}. 

Another interesting and important problem for further research would be  determination of other velocities like  the group velocity or the ratio-of-units velocity (see e.g. \cite{Blo77} or \cite{Gur01}) for the damped waves described by the fractional wave equations and comparison them each to other at least for the linear equations with the constant coefficients. Finally, fractional wave equations with non-constant coefficients as well as qualitative behavior of solutions of non-linear fractional wave equations would be worth to consider.

\vspace{0.3cm}

\noindent
{\bf Acknowledgment:} The author is thankful to Prof. Francesco Mainardi for useful and stimulating discussions regarding the subject of the paper during author's visit to the University of Bologna in December 2011.

\end{document}